# Fundamental limits for non-contact transfers between two bodies

Philippe Ben-Abdallah[1] and Karl Joulain[2]

1) Laboratoire de Thermocinétique, CNRS UMR 6607, Ecole Polytechnique de l'Université de Nantes, 44 306 Nantes cedex 03, France.

2) Institut P', CNRS-Université de Poitiers-CNRS UPR 3346, 86022 Poitiers Cedex, France.

**Abstract.** We investigate energy and momentum non-contact exchanges between two arbitrary flat media separated by a gap. This problem is revisited as a transmission problem of individual system eigenmodes weighted by a transmission probability obtained either from fluctuational electrodynamics or quantum field theory. An upper limit for energy and momentum flux is derived using a general variational approach. The corresponding optimal reflectivity coefficients are given both for identical and different media in interaction.

[1] Electronic mail : pba@univ-nantes.fr

[2] Electronic mail : karl.joulain@univ-poitiers.fr

Two arbitrary media in relative motion or at rest and separated by a gap continually exchange in permanence energy and momentum [1] throughout the thermally and quantum fluctuating electromagnetic field they radiate in their surrounding. At long separation distance compared to the Wien wavelength ($\ell \gg \lambda_T = c\hbar/(k_B T)$), energy radiative transfer is maximal when both media behave like blackbodies [2]. In this situation, the transfer is driven by the famous Stefan-Boltzmann law and exchanges only depends on the difference of media temperatures power four. When two bodies are in relative motion, the momentum exchanges through Doppler-shifted photons give rise to a van der Waals interaction [3-5] also called the van der Waals friction stress which is opposed to the motion. At long separation distances, this transfer is maximal when both media are perfect absorbers [5]. At subwavelength scale, the situation radically changes for both energy and momentum exchanges. Indeed, in this case, the presence of evanescent modes gives rise to wave phenomena such as tunnelling and interferences which drastically affect the transfers. Hence, energy and momentum exchanges can respectively exceed by several orders of magnitude the blackbody limit [6-10] for heat transfer and the van der Waals friction intensity between two perfect absorbers [4, 5, 11-13]. However, these results raised new questions. Are there fundamental limits for these exaltation mechanisms for energy [14] and momentum transfer between two bodies ? If they exist, what are these ultimate limits [15] and what are the media required to reach such values? In this paper we use calculus of variations principles in order to bring a general answer to these questions. In addition, we introduce a general Landauer-type formulation [16] for the transfer, to describe non-contact exchanges as a simple transmission problem. Relating this formulation to the results predicted by the fluctuational electrodynamic [17] and the quantum field [18] theory we derive energy and momentum transmission probabilities for both radiative and non-radiative photons. Moreover we derive and give a physical interpretation for the optimal conditions maximizing exchanges between two identical or distinct bodies. Finally, we examine near-field heat transfer between two media which support surface polaritons and compare it with the fundamental limits of transfer.

To start, let us consider two resonators A and B that we assume, for the sake of clarity, plane and layered, separated by a vacuum gap of thickness $\ell$. These resonators are reservoirs of radiative and non-radiative electromagnetic modes that are entirely defined by a couple $(\omega, \mathbf{q})$, $\omega$ being the mode angular frequency and $\mathbf{q}$ the parallel component of its wavevector. Each of these modes carries a quantum of energy $e = \hbar\omega$ and a quantum of momentum $m = \hbar|\mathbf{q}|$ (we henceforth denote generically $\zeta$ any of these quantities). Whatever the separation distance between the two reservoirs, a $\zeta$-flux is exchanged. This exchange occur through radiative modes coupling at long separation distances and both through non-radiative and radiative modes at short distances. If these reservoirs are maintained in nonequilibrium situations at two distinct temperatures $T_A$ and $T_B$ and are animated by a relative parallel motion with a velocity $\mathbf{v}_r$ then a certain amount of $\zeta$ can be transmitted from A to B throughout a coupling channel with a transmission probability $\tau_{A \to B}^{\zeta}(\omega, \mathbf{q}, \ell) \le 1$. Then, using a Landauer-type formulation for this $\zeta$-transfer [16] we have

$$\zeta_{A \to B} = \sum_{\omega, \mathbf{q}} f(\omega, T_A, \mathbf{v}_r, \mathbf{q}) \tau_{A \to B}^{\zeta}(\omega, \mathbf{q}, \ell) \qquad (1)$$

where f is a function which depends on the nature of exchanges we are dealing with. Suppose now that the transfer occur during a time $\Delta t$. Mode angular frequency are quantified as well as parallel mode if we suppose that the system is bounded by an arbitrary large square box area $S = L^2$. Then, we have

$$\omega = n_t \frac{\pi}{\Delta t} \text{ and } \mathbf{q} = n_x \frac{2\pi}{L} \mathbf{e}_x + n_y \frac{2\pi}{L} \mathbf{e}_y, \qquad (2)$$

where $(\mathbf{e}_x, \mathbf{e}_y)$ denote two orthogonal vectors in the plane parallel to the interfaces, $n_t$ is a positive integer, $n_x$ and $n_y$ are relative integers respectively. This quantification of mode specifies the number of channels available in the $(q,\omega)$-space. Examining the $\zeta$ exchange between A and B (resp. B and A) from this surface during a time interval $\Delta t$ we get, after transforming the discrete summation in (1) into an integration over a continuum

$$\zeta_{A \to B} = \frac{1}{2} \int_0^\infty \frac{d\omega}{(\pi/\Delta t)} \int \frac{d\mathbf{q}}{(2\pi/L)^2} f(\omega, T_A, \mathbf{v}_r, \mathbf{q}) \tau_{A \to B}^{\zeta}(\omega, \mathbf{q}, \ell). \qquad (3)$$

Note that this transformation from discrete to continuous summation is valid when the dimension L and time interval $\Delta t$ are macroscopic. It follows that the net flux of $\zeta$ exchanged between both media per unit surface reads

$$\varphi_{total}^{\zeta} = \frac{1}{8\pi^3} \int_0^{\infty} d\omega \int d\mathbf{q} \, [f(\omega, T_A, \mathbf{v}_r, \mathbf{q}) \tau_{A \to B}^{\zeta}(\omega, \mathbf{q}, \ell) - f(\omega, T_B, \mathbf{v}_r, \mathbf{q}) \tau_{B \to A}^{\zeta}(\omega, \mathbf{q}, \ell)] . \qquad (4)$$

At equilibrium $\varphi_{total}^{\zeta} = 0$ and $f(\omega, T_A, \mathbf{v}_r, \mathbf{q}) = f(\omega, T_B, \mathbf{v}_r, \mathbf{q})$ so that from (4) we have $\tau_{A \to B}^{\zeta} \equiv \tau_{B \to A}^{\zeta}$.

On the other hand, from fluctuational electrodynamics [1, 17] or quantum field [18] theory any $\zeta$-flux can be casted into a general form

$$\varphi_{total}^{\zeta} = \int_0^{\omega_{max}} d\omega F_{\zeta}(\omega, T_A, T_B, v_r) \phi(\omega) \qquad (5)$$

Let us separate propagative and evanescent modes contributions. We have $\phi(\omega) = \int_{q<\omega/c} d\mathbf{q} \, L_{prop}[R_A(\mathbf{q}), R_B(\mathbf{q}), \mathbf{q}]$ and $\phi(\omega) = \int_{q>\omega/c} d\mathbf{q} \, L_{eva}[R_A(\mathbf{q}), R_B(\mathbf{q}), \mathbf{q}]$. In this generic expression $F_{\zeta}$ depends on the type of transfer that is considered and $L_{prop}$ and $L_{eva}$ are functionals of the reflectivity $R_{A,B}$ of exchanging media. Due to the exponential damping of evanescent waves, q is limited, at a given frequency, to a value below the cutt-off wavector $q_c = \sqrt{4/\ell^2 + (\omega/c)^2}$. According to the fundamental lemma of calculus of variations, when the monochromatic flux is extremal, the real and imaginary parts of reflectivity coefficients satisfy the so called Euler-Lagrange's (EL) equations

$$\frac{\partial L_{prop,eva}}{\partial \operatorname{Re}[R_{A,B}]} = 0 \text{ and } \frac{\partial L_{prop,eva}}{\partial \operatorname{Im}[R_{A,B}]} = 0 \qquad (6)$$

Now let us consider the transfer of heat between two bodies in non-equilibrium thermal situation. According to the fluctuational electrodynamic theory [1, 6-7] and using the azimuthal symmetry of problem (i.e. $\int d\mathbf{q} = 2\pi \int q dq$) we can see that $\omega_{max} = \infty$, $L_{prop} = \frac{1}{4\pi^2} q \frac{(1-|R_A|^2)(1-|R_B|^2)}{|1 - R_A R_B e^{-2i\gamma\ell}|^2}$,

$$L_{eva} = \frac{1}{\pi^2} q \frac{\text{Im}(R_A)\text{Im}(R_B)}{\left|1 - R_A R_B e^{-2\gamma''\ell}\right|^2} e^{-2\gamma''\ell} \quad \text{and} \quad F = \Theta(\omega,T_A) - \Theta(\omega,T_B), \quad \Theta(\omega,T) \equiv \hbar\omega/[\exp(\hbar\omega/k_B T) - 1]$$

being the mean energy of a Planck oscillator at equilibrium and $\gamma = \sqrt{(\omega/c)^2 - q^2}$ (with $\text{Im}\,\gamma = \gamma'' \geq 0$) the normal component of the wave vector in the intracavity space. Then, given the reflectivity of one of two interacting media, let say $R_B$, the EL's equations (6) lead, after a straightforward calculation, to the solution in the non-radiative range (i.e. $q > \omega/c$)

$$R_A^{opt} = e^{i\text{Arg}(R_B)} e^{2\gamma''\ell} \tag{7}$$

while in the particular case of identical media

$$R_A^{opt} = e^{i\varphi} e^{2\gamma''\ell} \text{ for any } 0 < \varphi < \pi/2. \tag{8}$$

Note that for both geometrical configurations $R_A^{opt} = 0$ is the optimal reflectivity in the radiative range. It follows from these expressions that $\dfrac{(1 - |R_A^{opt}|^2)(1 - |R_B|^2)}{\left|1 - R_A^{opt} R_B e^{-2i\gamma\ell}\right|^2} = 1$ and

$\dfrac{\text{Im}(R_A^{opt})\text{Im}(R_B)}{\left|1 - R_A^{opt} R_B e^{-2\gamma''\ell}\right|^2} e^{-2\gamma''\ell} = \dfrac{1}{4}$. Therefore, by identifying (4) and (5) we obtain

$$\tau_{A \to B}^e(\omega, q, \ell) = \begin{cases} \dfrac{(1 - |R_A|^2)(1 - |R_B|^2)}{\left|1 - R_A R_B e^{-2i\gamma\ell}\right|^2}, & q < \omega/c \\[2ex] 4\dfrac{\text{Im}(R_A)\text{Im}(R_B)}{\left|1 - R_A R_B e^{-2\gamma''\ell}\right|^2} e^{-2\gamma''\ell}, & q > \omega/c \end{cases} \quad \text{and} \quad f(\omega, T_A, v_r, q) = \frac{1}{4\pi^2} qF \text{ for any } q. \tag{9}$$

For $q < \omega/c$, $\tau_{A \to B}^e = 1$ corresponds to an energy exchange between two blackbodies that is a transfer of radiative waves between two perfect emitters. In this case, by performing integration of flux over all the spectrum it is direct to see from (4), after summation over the two polarizations states, that we recover the Stefan-Boltzmann law [2] (i.e. $\varphi_{A \to B}^e = \sigma(T_A^4 - T_B^4)$). When $q > \omega/c$, the condition $\tau_{A \to B}^e = 1$ corresponds to a perfect tunnelling of non-radiative photons. Also, we see from (4) that this transfer is maximal at a given frequency when the number of coupled (evanescent) modes per unit surface $N(\omega) \equiv \int_{\omega/c}^{q_c} \frac{q}{4\pi^2} \tau_{A \to B}^e(\omega, q) dq$ becomes maximum. This precisely occurs when $\tau_{A \to B}^e = 1$. In this case $N(\omega) = N_{max} = \frac{1}{2\pi^2 \ell^2}$ so that, by taking into account the two polarization states of non-radiative photons, the upper limit for the near-field heat transfer between two media reads

$$\varphi_{A \to B}^{e,max} = 2N_{max} \int_0^\infty d\omega [\Theta(\omega, T_A) - \Theta(\omega, T_B)] = \frac{k_B^2}{6\hbar \ell^2}[T_A^2 - T_B^2]. \qquad (10)$$

If we assume that $T_A = T_B + \delta T$ with $\delta T / T_A \ll 1$ we can introduce a heat transfer coefficient from $\varphi_{A \to B}^e = h_e \delta T$. From (10) after linearization, we see that $h_e^{max} = \frac{2g_0}{\pi \ell^2}$ where $g_0 = \pi^2 k_B^2 T_A / 3h$ is the quantum of thermal conductance at $T_A$. Note that the separation distance $\ell$ cannot go below the scales where non local effects appear. Thus, for metals it is limited by the Thomas-Fermi screening length while for dielectrics it is the interatomic distance $a$ which defines the lower limit [19]. For silicon, interatomic distance is roughly 0.24 nm so that the ultimate conductance at 300 K between two silicon samples is approximately $3.10^{10}$ W.m$^{-2}$.K$^{-1}$ whereas typical conductance between bulk silicon atomic layers is $\lambda_{Si}/a = 6.10^{11}$ W.m$^{-2}$.K$^{-1}$, $\lambda_{Si}$ being the thermal conductivity of silicon. Note that if non-contact heat transfer can reach values several orders of magnitude larger than what is exchanged in far-field (conductance of $4\sigma T^3 = 6$ W.m$^{-2}$.K$^{-1}$ at 300 K), we remark that it can hardly beat the classical thermal conduction of bulk materials.

If we turn out to the non-contact friction problem at zero temperature between two reservoirs in parallel relative motion at non-relativistic velocity $v_r$, then according to the quantum field theory [4,17] $\omega_{max} = q_x v_r$, $L_{prop} \approx 0$ and $L_{eva} = \frac{\hbar}{4\pi^3} q_x \frac{\text{Im}(R_A) \text{Im}(R_B^-)}{|1 - R_A R_B^- e^{-2q\ell}|^2} e^{-2q\ell}$, where $R_B^- = R_B(\omega - q_x v_r)$ and $q = \sqrt{q_x^2 + q_y^2}$. In that case, frictional stress optimization

$$\varphi_{A \to B}^{m} = \int_{-\infty}^{\infty} dq_y \int_{0}^{\infty} dq_x \int_{0}^{q_x v_r} d\omega L_{eva}$$ leads by analogy with the heat transfer problem to

$R_{A}^{opt} = e^{i\text{Arg}(R_{B}^{-})} e^{2q\ell}$ so that

$$\tau_{A \to B}^{m}(\omega, q, \ell) = 4 \frac{\text{Im}(R_A)\text{Im}(R_B^{-})}{\left|1 - R_A R_B^{-} e^{-2q\ell}\right|^2} e^{-2q\ell} \text{ and } f(\omega, T_A, v_r, \mathbf{q}) = \frac{\hbar}{16\pi^3} q_x. \quad (11)$$

It follows from (5) that the upper limit for the frictional stress between two media in relative motion reads

$$\varphi_{A \to B}^{m,\max} = \frac{\hbar}{8\pi^3} v_r \int_{0}^{2/\ell} dq\, q^3 \int_{-\pi/2}^{\pi/2} d\theta \cos^2\theta = \frac{2\hbar}{\pi^2 \ell^4} v_r. \quad (12)$$

Now let us compare energy we can exchange between two semi-infinite media with the ultimate values predicted above. Assuming these media are identical with a dielectric permittivity given by a Lorentz-Drude model $\varepsilon(\omega) = 1 + \frac{2(\Omega^2 - \omega_0^2)}{\omega_0^2 - \omega(i\Gamma + \omega)}$ where $\Omega$ and $\omega_0$ denote a longitudinal and transversal like optical phonon pulsations while $\Gamma$ is a damping factor. These media support a surface polariton at $\omega = \Omega$. Close to this mode $R_A \approx \frac{\varepsilon - 1}{\varepsilon + 1}$ so that from (9)

$$\tau_{A \to B}^{e} = \frac{4(\Omega^2 - \omega_0^2)^2 \Gamma^2 \omega^2}{[(\Omega^2 - \omega^2)^2 + \Gamma^2 \omega^2]^2 e^{2x} - 2[(\Omega^2 - \omega^2)^2 - \Gamma^2 \omega^2](\Omega^2 - \omega_0^2)^2 + (\Omega^2 - \omega_0^2)^4 e^{-2x}} \quad (13)$$

with $x = \ell\sqrt{q^2 - (\omega/c)^2}$. From this formula, we see that the condition $\tau_{A \to B}^{e} = 1$ corresponds to a curve in the $(\omega, q)$ plane defined by $e^{2x} = \frac{(\Omega^2 - \omega_0^2)^2}{(\Omega^2 - \omega^2)^2 + \Gamma^2 \omega^2}$. In the neighbourhood of this curve the transfer of heat by tunnelling is very efficient. At the surface polariton frequency and far from the light line $(q \gg \omega/c)$, $\tau_{A \to B}^{e}$ is maximum on this curve until $q_{\max} \sim \ln(\Omega/\Gamma)$ and $\tau_{A \to B}^{e} > 1/2$ around this curve in a domain of typical width $\Delta q = 2/\ell$ in wavevector and $\Delta\omega = \sqrt{2}\Gamma$ in angular frequency as shown in Fig.1. In this figure, note that $q = 1/\ell$ corresponds to $qc/\omega = 20$ so that the region where $\tau_{A \to B}^{e} = 1$ goes to high q values and contributes to high heat transfer values. Comparing the ratio between the heat transfer coefficient $h_e$ due to polaritons and the ultimate value $h_e^{max}$ calculated from above we show that

$$h_e / h_e^{max} \approx \frac{\ln(\Omega/\Gamma)}{(\Omega/\Gamma)} \left(\frac{\hbar\Omega}{k_B T}\right)^3 \frac{e^{\hbar\Omega/k_B T}}{(e^{\hbar\Omega/k_B T} - 1)^2} \quad . \tag{14}$$

It appears from this expression and from numerical calculation of heat transfer at T=300K when $\Omega = 6 \times 10^{13}$ rad.s$^{-1}$ and and $\Gamma = 1.5 \times 10^{13}$ rad.s$^{-1}$ that it reaches about 25% of the ultimate heat transfer. Polaritons are thus excellent candidates to maximize non contact heat transfer between materials where they could be used in next generation thermophotovoltaic devices to enhance conversion of radiation into electricity. These results are in full agreement with the works of Wang et al. [20] on the magnification of transfers between two optimized dielectrics.

We have theoretically derived the fundamental limits, for energy and momentum exchanges between two parallel layered media separated by a vacuum gap at rest in nonequilibrium thermal situation and in relative motion at zero temperature respectively. The corresponding optimal reflection coefficients have been precisely determined using basic principles of calculus of variations. Our approach is general provided that non-local effects can be neglected and the sliding velocity is non-relativistic. It could be extended to optimize exchanges between any structure shapes using the concept of generalized scattering operators rather than that of reflectivities. Quantum friction at finite temperature could also be investigated in the same way. In addition, a Landauer-type formulation of non-contact exchanges has been presented both for radiative and non-radiative modes. We have shown that while the near-field heat transfer at a given frequency is maximum when the number of coupled evanescent modes per unit surface is maximum, the momentum transfer mediated by shearing depends on the energy density of electromagnetic modes on the surfaces in interaction. These results should provide a guidance for the design of composite materials dedicated to exalted transfers in near-field technology.

**Captions list**

Fig. 1 Transmission probability in p-polarization of quantum of heat between two (massive) samples separated by a distance $\ell = 100$ nm with Drude-Lorentz parameters $\Omega = 6 \times 10^{13}$ rad.s$^{-1}$ and $\Gamma = 1.5 \times 10^{13}$ rad.s$^{-1}$.

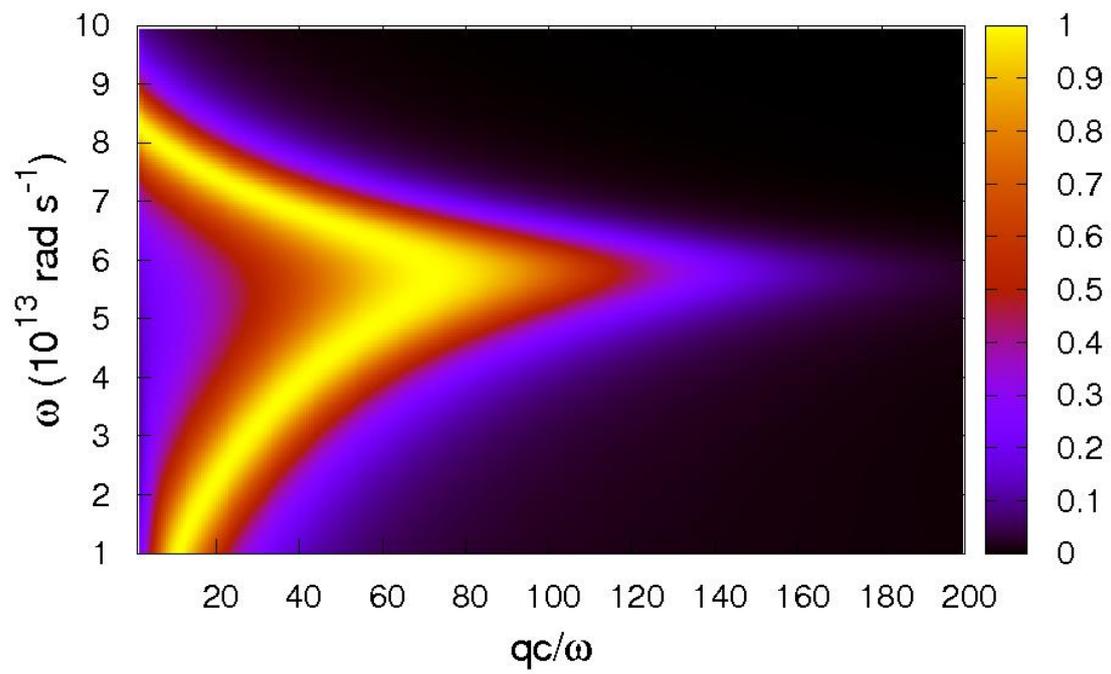

**Figure 1**